\documentclass[letterpaper, 12pt]{article}[2000/05/19]
\usepackage[english]{babel}
\usepackage{amsfonts,amsmath,amssymb,amsthm,latexsym,amscd,mathrsfs}
\usepackage{ifthen,cite}
\usepackage[bookmarksnumbered=true]{hyperref}
\hypersetup{pdfpagetransition={Split}}

\newcommand{\evenhead}{Author \ name}
\newcommand{\oddhead}{Article \ name}
\newcommand{\theArticleName}{Article \ name}

\newcommand{\FirstPageHeading}[1]{\thispagestyle{empty}%
\noindent\raisebox{0pt}[0pt][0pt]{\makebox[\textwidth]{\protect\footnotesize \sf }}\par}

\newcommand{\ArticleName}[1]{\renewcommand{\theArticleName}{#1}\vspace{-2mm}\par\noindent {\LARGE\bf  #1\par}}
\newcommand{\Author}[1]{\vspace{5mm}\par\noindent {\Large  #1\par} \par\vspace{2mm}\par}
\newcommand{\Address}[1]{\vspace{2mm}\par\noindent {\it #1} \par}
\newcommand{\Email}[1]{\ifthenelse{\equal{#1}{}}{}{\par\noindent {\rm E-mail: }{\it  #1} \par}}
\newcommand{\URLaddress}[1]{\ifthenelse{\equal{#1}{}}{}{\par\noindent {\rm URL: }{\tt  #1} \par}}
\newcommand{\EmailD}[1]{\ifthenelse{\equal{#1}{}}{}{\par\noindent {$\phantom{\dag}$~\rm E-mail: }{\it  #1} \par}}
\newcommand{\URLaddressD}[1]{\ifthenelse{\equal{#1}{}}{}{\par\noindent {$\phantom{\dag}$~\rm URL: }{\tt  #1} \par}}

\newcommand{\Abstract}[1]{\vspace{6mm}\par\noindent\hspace*{10mm}
\parbox{140mm}{\small {\bf Abstract.} #1}\par}
\newcommand{\Keywords}[1]{\vspace{3mm}\par\noindent\hspace*{10mm}
\parbox{140mm}{\small {\bf Key words:} \rm #1}\par}
\newcommand{\Classification}[1]{\vspace{3mm}\par\noindent\hspace*{10mm}
\parbox{140mm}{\small {\it 2010 Mathematics Subject Classification:} \rm #1}\vspace{3mm}\par}
\newcommand{\ShortArticleName}[1]{\renewcommand{\oddhead}{#1}}
\newcommand{\AuthorNameForHeading}[1]{\renewcommand{\evenhead}{#1}}

\long\def\@makecaption#1#2{
  \sbox\@tempboxa{\small \textbf{#1.}\ \ #2}%
  \ifdim \wd\@tempboxa >\hsize
    {\small \textbf{#1.}\ \ #2}\par \else
    \global \@minipagefalse
    \hb@xt@\hsize{\hfil\box\@tempboxa\hfil}%
  \fi \vskip\belowcaptionskip}

\def\numberwithin#1#2{\@ifundefined{c@#1}{\@nocounterr{#1}}{%
  \@ifundefined{c@#2}{\@nocnterr{#2}}{%
  \@addtoreset{#1}{#2}%
  \toks@\@xp\@xp\@xp{\csname the#1\endcsname}%
  \@xp\xdef\csname the#1\endcsname
    {\@xp\@nx\csname the#2\endcsname.\the\toks@}}}}
\def\E^#1{{\buildrel #1 \over\vee}}

{\theoremstyle{definition}

}

\begin{document}

\FirstPageHeading{V.I. Gerasimenko}

\ShortArticleName{Enskog equation for granular gases}

\AuthorNameForHeading{M.S. Borovchenkova, V.I. Gerasimenko}

\ArticleName{On the non-Markovian Enskog Equation\\ for Granular Gases}

\Author{M.S. Borovchenkova$^\ast$\footnote{E-mail: \emph{borovchenkova@ukr.net}}
        and V.I. Gerasimenko $^\ast$$^\ast$\footnote{E-mail: \emph{gerasym@imath.kiev.ua}}}

\Address{$^\ast$\hspace*{2mm}Taras Shevchenko National University of Kyiv,\\
    \hspace*{4mm}Department of Mechanics and Mathematics,\\
    \hspace*{4mm}2, Academician Glushkov Av.,\\
    \hspace*{4mm}03187, Kyiv, Ukraine}

\Address{$^\ast$$^\ast$Institute of Mathematics of NAS of Ukraine,\\
    \hspace*{4mm}3, Tereshchenkivs'ka Str.,\\
    \hspace*{4mm}01601, Kyiv-4, Ukraine}

\bigskip

\Abstract{We develop a rigorous formalism for the description of the kinetic
evolution of many-particle systems with the dissipative interaction. The
relationships of the evolution of a hard sphere system with inelastic collisions
described within the framework of marginal observables governed by the dual BBGKY
hierarchy and the evolution of states described by the Cauchy problem of the Enskog
kinetic equation for granular gases are established. Moreover, we consider
the Boltzmann--Grad asymptotic behavior of the constructed non-Markovian
Enskog kinetic equation for granular gases in a one-dimensional space.}

\bigskip

\vspace{2pc}
\Classification{82C05; 82C40; 82D99; 35Q20; 35Q82}

\makeatletter
\renewcommand{\@evenhead}{
\hspace*{-3pt}\raisebox{-7pt}[\headheight][0pt]{\vbox{\hbox to \textwidth {\thepage \hfil \evenhead}\vskip4pt \hrule}}}
\renewcommand{\@oddhead}{
\hspace*{-3pt}\raisebox{-7pt}[\headheight][0pt]{\vbox{\hbox to \textwidth {\oddhead \hfil \thepage}\vskip4pt\hrule}}}
\renewcommand{\@evenfoot}{}
\renewcommand{\@oddfoot}{}
\makeatother

\newpage
\vphantom{math}

\protect\tableofcontents
\vspace{0.5cm}

\section{Introduction}
A granular gas is a dynamical system of significant interest not only in view of its
applications but also as a many-particle system displaying a collective behavior that
differs from the statistical behavior of usual gases; for example, it is related to
typical macroscopic properties \cite{C3}-\cite{M09}.

As is known, the collective behavior of many-particle systems can be effectively described
within the framework of a one-particle marginal distribution function governed by the
kinetic equation in a suitable scaling limit of underlying dynamics \cite{GH}-\cite{CGP97}.
At present, the considerable advances are being in the rigorous derivation of the Boltzmann kinetic
equation of a system of hard spheres in the Boltzmann--Grad scaling limit
\cite{PG1}-\cite{PSS}. At the same time, many recent papers \cite{VS}-\cite{T05} (and
see references therein) are  considering the Boltzmann-type and the Enskog-type kinetic equations for
inelastically interacting hard spheres, modelling granular gases, as the
original evolution equations and the rigorous derivation of such kinetic equations remain
a problem \cite{BDS}-\cite{P08}.

The goal of this paper is to develop an approach based on the dynamics of particles
with the dissipative interaction to properly justify the kinetic equations that
previous works have already applied a priori to the description of granular gases.
In the paper, we consider the problem of potentialities inherent in the description
of the evolution of states of a hard sphere system with inelastic collisions in terms
of a one-particle distribution function. We established that, in fact, if the initial
state is completely specified by a one-particle marginal distribution function, then
all possible states at an arbitrary moment of time can be described within the framework
of a one-particle distribution function without any approximations.

To outline the structure of the paper and the main results, in section 2,
we develop an approach to the description of the kinetic evolution of hard spheres
with inelastic collisions within the framework of the evolution of marginal observables.
Then, in section 3, the main results related to the origin of the kinetic evolution of
granular gases are stated. We prove that underlaying dynamics governed by the dual BBGKY
hierarchy for marginal observables can be completely described within the framework of
the one-particle marginal distribution function governed by the non-Markovian Enskog
kinetic equation with inelastic collisions. In this case, we prove that all possible
correlations, creating by hard sphere dynamics, are described by the explicitly defined
marginal functionals with respect to the solution of the established kinetic equation.
In section 4, we consider a one-dimensional granular gas. The Boltzmann--Grad asymptotic
behavior of the constructed non-Markovian Enskog kinetic equation with
inelastic collisions in a one-dimensional space is outlined. Finally, in section 5, we
conclude with some observations and perspectives for future research.

\section{Hierarchies of evolution equations for granular gases}
It is well known that many-particle systems are descried in terms of two sets of objects:
observables and states. The functional of the mean value of observables defines a duality
between observables and states, and as a consequence two approaches to the
description of the evolution exist. Usually, the evolution of many-particle systems is described
within the framework of the evolution of states by the BBGKY hierarchy for marginal
distribution functions. An equivalent approach to this description is given in terms of
the marginal observables governed by the dual
BBGKY hierarchy. In the same framework, the systems of particles with the dissipative
interaction, namely hard spheres with inelastic collisions, can be described.

\subsection{The dual BBGKY hierarchy and semigroups of operators of hard spheres with inelastic collisions}
We consider a system of a non-fixed, i.e. arbitrary, but finite average number of identical
particles of a unit mass, interacting as hard spheres with inelastic collisions. Every hard
sphere with the diameter $\sigma>0$ is characterized by the phase coordinates
$(q_{i},p_{i})\equiv x_{i}\in\mathbb{R}^{3}\times\mathbb{R}^{3},\,i\geq1.$

Let $C_{\gamma}$ be the space of sequences $b=(b_0,b_1,\ldots,b_n,\ldots)$ of bounded
continuous functions $b_n\in C_n$ defined on the phase space of $n$ hard spheres that
are symmetric with respect to the permutations of the arguments $x_1,\ldots,x_n$, equal
to zero on the set of forbidden configurations $\mathbb{W}_n\doteq\big\{(q_1,\ldots,q_n)
\in\mathbb{R}^{3n}\big||q_i-q_j|<\sigma$ for at least one pair $(i,j): i\neq j\in(1,\ldots,n)\big\}$
and equipped with the norm
$\|b\|_{C_{\gamma}}=\max_{n\geq 0}\,\frac{\gamma^{n}}{n!}\,\|b_n\|_{C_n}=\max_{n\geq 0}\,
\frac{\gamma^{n}}{n!}\,\sup_{x_1,\ldots,x_n}|b_n(x_1,\ldots,x_n)|$. We denote the set of
continuously differentiable functions with compact supports by $C_{n,0}\subset C_n$.

Within the framework of observables, the evolution of a system of an arbitrary but finite
average number of hard spheres is described by the sequences
$B(t)=(B_0,B_{1}(t,x_1),\ldots,B_{s}(t,x_{1},\ldots,x_{s}),$ $\ldots)\in C_{\gamma}$
of the marginal ($s$-particle) observables $B_s(t,x_1,\ldots,x_s)$ defined on the phase
space of $s\geq1$ hard spheres that are symmetric with respect to the permutations of
the arguments $x_1,\ldots,x_n$, equal to zero on the set $\mathbb{W}_s$, and for $t\geq0$
they are governed by the Cauchy problem of the weak formulation of the
dual BBGKY hierarchy \cite{BGer}
\begin{eqnarray}\label{dh}
   &&\hskip-9mm \frac{\partial}{\partial t}B_{s}(t,x_1,\ldots,x_s)=\big(\sum\limits_{j=1}^{s}\mathcal{L}(j)B_{s}(t)+
      \sum\limits_{j_1<j_{2}=1}^{s}\mathcal{L}_{\mathrm{int}}(j_1,j_{2})B_{s}(t)\big)(x_1,\ldots,x_s)+\\
   &&+\sum_{j_1\neq j_{2}=1}^s
      \big(\mathcal{L}_{\mathrm{int}}(j_1,j_{2})B_{s-1}(t)\big)(x_1,\ldots,x_{j_1-1},x_{j_1+1},\ldots,x_s),\nonumber\\
      \nonumber\\
  \label{dhi}
   &&\hskip-9mm B_{s}(t,x_1,\ldots,x_s)\mid_{t=0}=B_{s}^0(x_1,\ldots,x_s),\quad s\geq1,
\end{eqnarray}
where on the set $C_{s,0}\subset C_s$ the free motion Liouville operator $\mathcal{L}(j)$ and
the operator of inelastic collisions $\mathcal{L}_{\mathrm{int}}(j_1,j_{2})$ are defined by
the following formulas
\begin{eqnarray}\label{Lj}
   &&\mathcal{L}(j)\doteq\langle p_j,\frac{\partial}{\partial q_j}\rangle,
\end{eqnarray}
and
\begin{eqnarray}\label{Lint}
    &&\hskip-18mm\mathcal{L}_{\mathrm{int}}(j_{1},j_{2})b_{n}\doteq
        \sigma^2\int_{\mathbb{S}_{+}^2}d\eta\langle\eta,(p_{j_{1}}-p_{j_{2}})\rangle
        \big(b_n(x_1,\ldots,p_{j_{1}}^\ast,q_{j_{1}},\ldots,p_{j_{2}}^\ast,q_{j_{2}},\ldots,x_n)-\\
    &&-b_n(x_1,\ldots,x_n)\big)\delta(q_{j_{1}}-q_{j_{2}}+\sigma\eta),\nonumber
\end{eqnarray}
respectively. In (\ref{Lj}),(\ref{Lint}) the following notations are used: the symbol
$\langle\cdot,\cdot\rangle$ means a scalar product, $\delta$ is the Dirac measure,
$\mathbb{S}_{+}^{2}\doteq\{\eta\in\mathbb{R}^{3}\big|\left|\eta\right|=1,\,
\langle\eta,(p_{j_{1}}-p_{j_{2}})\rangle\geq0\}$ and the post-collision momenta are
determined by
\begin{eqnarray}\label{col}
     &&p_{j_{1}}^\ast=p_{j_{1}}-(1-\varepsilon)\,\eta\langle\eta,(p_{j_{1}}-p_{j_{2}})\rangle,\\
     &&p_{j_{2}}^\ast=p_{j_{2}}+(1-\varepsilon)\,\eta\langle\eta,(p_{j_{1}}-p_{j_{2}})\rangle,\nonumber
\end{eqnarray}
where $\varepsilon=\frac{1-e}{2}\in [0,\frac{1}{2})$ and $e\in(0,1]$ is a restitution
coefficient \cite{GG04}.

We give explicit examples of recurrence evolution equations (\ref{dh}):
\begin{eqnarray*}
   &&\hskip-9mm \frac{\partial}{\partial t}B_{1}(t,q_1,p_1)=
        \langle p_1,\frac{\partial}{\partial q_1}\rangle B_{1}(t,q_1,p_1),\\
   &&\hskip-9mm \frac{\partial}{\partial t}B_{2}(t,x_1,x_2)=
        \sum\limits_{j=1}^{2}\langle p_j,\frac{\partial}{\partial q_j}\rangle B_{2}(t,x_1,x_2)+\\
   &&\hskip-5mm +\sigma^2\int_{\mathbb{S}_{+}^2}d\eta\langle\eta,(p_{1}-p_{2})\rangle
        \big(B_{2}(t,x_{1}^\ast,x_{2}^\ast)-B_{2}(t,x_1,x_2)\big)\delta(q_{1}-q_{2}+\sigma\eta)+\\
   &&\hskip-5mm +\sigma^2\int_{\mathbb{S}_{+}^2}d\eta\langle\eta,(p_{1}-p_{2})\rangle
        \big(B_{1}(t,x_{1}^\ast)-B_{1}(t,x_1)\big)\delta(q_{1}-q_{2}+\sigma\eta)+\\
   &&\hskip-5mm +\sigma^2\int_{\mathbb{S}_{+}^2}d\eta\langle\eta,(p_{1}-p_{2})\rangle
        \big(B_{1}(t,x_{2}^\ast)-B_{2}(t,x_2)\big)\delta(q_{1}-q_{2}+\sigma\eta).
\end{eqnarray*}

We refer to recurrence evolution equations (\ref{dh}) as the dual BBGKY hierarchy
for hard spheres with inelastic collisions or for granular gases.

The first term of a generator of the dual BBGKY hierarchy (\ref{dh}) is the Liouville operator
$\mathcal{L}_s=\sum_{j=1}^{s}\mathcal{L}(j)+\sum_{j_1<j_{2}=1}^{s}\mathcal{L}_{\mathrm{int}}(j_1,j_{2})$,
which is an infinitesimal generator of the semigroup of operators $S_{s}(t),\,t\geq0,$
of a system of $s$ inelastically interacting hard spheres. The semigroup of operators
$S_{s}(t),\,t\geq0,$ is defined almost everywhere on the phase space
$\mathbb{R}^{3s}\times(\mathbb{R}^{3s}\setminus\mathbb{W}_s)$, namely, outside the set
$\mathbb{M}_{s}^0$ of the Lebesgue measure zero, as a shift operator of the arguments
of functions from the space $C_s$ along the phase trajectories of $s$ hard spheres with
inelastic collisions
\begin{eqnarray}\label{Sspher}
  &&\hskip-5mm (S_{s}(t)b_{s})(x_{1},\ldots,x_{s})\equiv
      S_{s}(t,1,\ldots,s)b_{s}(x_{1},\ldots,x_{s})\doteq \\
  &&\hskip+5mm \doteq\begin{cases}
         b_{s}(X_{1}(t),\ldots,X_{s}(t)),
         \hskip+5mm\mathrm{if}\,(x_{1},\ldots,x_{s})
         \in(\mathbb{R}^{3s}\times(\mathbb{R}^{3s}\setminus\mathbb{W}_{s})),\\
         0, \hskip+38mm \mathrm{if}\,(q_{1},\ldots,q_{s})\in\mathbb{W}_{s},
         \end{cases}\nonumber
\end{eqnarray}
where the function $X_{i}(t)\equiv X_{i}(t,x_{1},\ldots,x_{s})$ is a phase trajectory
of the $ith$ hard sphere constructed in \cite{P08} and the set $\mathbb{M}_{s}^0$ consists
from phase space points specified in the initial data $x_{1},\ldots,x_{s}$ that generate multiple
collisions during the evolution \cite{CGP97}.

On the space $C_s$, one-parameter mapping (\ref{Sspher}) is a bounded $\ast$-weak continuous
semigroup of operators \cite{DauL}, and $\big\|S_s(t)\big\|<1$.

Let $L^{1}_{\alpha}=\oplus^{\infty}_{n=0}\alpha^n L^{1}_{n}$ be the space of sequences
$f=(f_0,f_1,\ldots,f_n,\ldots)$ of integrable functions $f_n(x_1,\ldots,x_n)$ defined
on the phase space $\mathbb{R}^{3n}\times(\mathbb{R}^{3n}\setminus\mathbb{W}_n)$ of $n$
hard spheres that are symmetric with respect to the permutations of the arguments $x_1,\ldots,x_n$,
equal to zero on the set of forbidden configurations $\mathbb{W}_n$, and equipped with the norm
$\|f\|_{L^{1}_{\alpha}}=\sum_{n=0}^{\infty}\alpha^n \int dx_1\ldots dx_n|f_n(x_1,\ldots,$ $x_n)|$,
where $\alpha>1$ is a real number. We denote by $L_{0}^1\subset L^{1}_{\alpha}$ the everywhere
dense set in $L^{1}_{\alpha}$ of finite sequences of continuously differentiable functions with
compact supports.

On the space of integrable functions $L^{1}_{s}$, the semigroup of operators
$S_{s}^\ast(t),\,t\geq0,$ is defined adjoint to the semigroup of operators (\ref{Sspher}) in the sense
of the continuous linear functional (the functional of mean values of observables)
\begin{eqnarray*}\label{av}
   &&\hskip-5mm\big\langle b\big|f\big\rangle=\sum\limits_{s=0}^{\infty}\,
      \frac{1}{s!}\int_{(\mathbb{R}^{3}\times\mathbb{R}^{3})^{s}}
      dx_{1}\ldots dx_{s}\,b_{s}(x_1,\ldots,x_s)f_{s}(x_1,\ldots,x_s).
\end{eqnarray*}
The adjoint semigroup of operators is defined by the Duhamel equation
\begin{eqnarray}\label{DuamN_1}
    &&\hskip-12mmS_s^\ast(t,1,\ldots,s)=\prod\limits_{i=1}^{s}S_1^\ast(t,i)+
       \int\limits_0^t d\tau \prod\limits_{i=1}^{s}S_{1}^\ast(t-\tau,i)
       \sum\limits_{j_{1}<j_{2}=1}^{s}\mathcal{L}_{\mathrm{int}}^\ast(j_{1},j_{2})
       S_{s}^\ast(\tau,1,\ldots,s),
\end{eqnarray}
where the operator $\mathcal{L}_{\mathrm{int}}^\ast(j_{1},j_{2})$ is given
by the formula
\begin{eqnarray}\label{aLint}
     &&\hskip-18mm\mathcal{L}_{\mathrm{int}}^\ast(j_{1},j_{2})f_{s}
        \doteq\sigma^2\int_{\mathbb{S}_{+}^2}d\eta\langle\eta,(p_{j_{1}}-p_{j_{2}})\rangle \big(\frac{1}{(1-2\varepsilon)^{2}}\,f_n(x_1,\ldots,p_{j_{1}}^\diamond,q_{j_{1}},\ldots,\\
     &&\hskip-8mm p_{j_{2}}^\diamond,q_{j_{2}},\ldots,x_s)\delta(q_{j_{1}}-q_{j_{2}}+\sigma\eta)-
        f_n(x_1,\ldots,x_s)\delta(q_{j_{1}}-q_{j_{2}}-\sigma\eta)\big),\nonumber
\end{eqnarray}
In expression (\ref{aLint}), the notations similar to formula (\ref{Lint}) are used, and the pre-collision
momenta, i.e. solutions of equations (\ref{col}), are determined as follows:
\begin{eqnarray}\label{scol}
     &&p_{j_{1}}^{\diamond}=p_{j_{1}}-
         \frac{1-\varepsilon}{1-2\varepsilon}\,\eta\langle\eta,(p_{j_{1}}-p_{j_{2}})\rangle,\\
     &&p_{j_{2}}^{\diamond}=p_{j_{2}}+
         \frac{1-\varepsilon}{1-2\varepsilon}\,\eta\langle\eta,(p_{j_{1}}-p_{j_{2}})\rangle.\nonumber
\end{eqnarray}
Hence, an infinitesimal generator of the semigroup of operators adjoint to semigroup (\ref{Lint})
is defined on $L^1_{0,s}$ as the operator: $\mathcal{L}_{s}^\ast=\sum_{j=1}^{s}\mathcal{L}^\ast(j)+
\sum_{j_{1}<j_{2}=1}^{s}\mathcal{L}_{\mathrm{int}}^\ast(j_{1},j_{2})$,
where we introduce the operator adjoint to free motion operator (\ref{Lj}):
$\mathcal{L}^{\ast}(j)\doteq-\langle p_j,\frac{\partial}{\partial q_j}\rangle$
and the operator $\mathcal{L}_{\mathrm{int}}^\ast(j_{1},j_{2})$ is defined by formula (\ref{aLint}).

On the space $L_n^1$ the one-parameter mapping defined by equation (\ref{DuamN_1}) is a bounded
strong continuous semigroup of operators.

We remark that, according to the definition of the marginal observables \cite{BG}
\begin{eqnarray*}\label{mo}
      &&\hskip-5mmB_{s}(t,x_1,\ldots,x_s)\doteq\sum_{n=0}^s\,\frac{(-1)^n}{n!}
           \sum_{j_1\neq\ldots\neq j_{n}=1}^s
           (S_{s-n}(t)A_{s-n}^0)((x_1,\ldots,x_s)\setminus(x_{j_1},\ldots,x_{j_{n}})),\quad s\geq 1,
\end{eqnarray*}
in terms of a sequence of the observables
$A^0=(A_0,A_{1}^0(x_1),\ldots,A_{s}^0(x_{1},\ldots,x_{s}),\ldots)\in C_{\gamma}$, the
dual BBGKY hierarchy (\ref{dh}) for hard spheres with inelastic collisions can be rigorously
derived due to the properties of semigroups (\ref{Sspher}) of a hard sphere system with
inelastic collisions.

\subsection{A solution expansion of the dual BBGKY hierarchy for granular gases}
On the space $C_{\gamma}$ for abstract initial-value problem (\ref{dh}),(\ref{dhi}),
the following statement is true.

A solution $B(t)=(B_{0},B_{1}(t,x_1),\ldots,B_{s}(t,x_1,\ldots,x_s),\ldots)$ of the Cauchy problem
(\ref{dh}),(\ref{dhi}) is determined by the expansions \cite{BG}
\begin{eqnarray}\label{sdh}
   &&\hskip-8mm B_{s}(t,x_1,\ldots,x_s)=\sum_{n=0}^s\,\frac{1}{n!}\sum_{j_1\neq\ldots\neq j_{n}=1}^s
      \mathfrak{A}_{1+n}\big(t,\{Y\setminus Z\},Z\big)\,B_{s-n}^0(x_1,\ldots,\\
   &&\hskip+25mm x_{j_1-1},x_{j_1+1},\ldots,x_{j_n-1},x_{j_n+1},\ldots,x_s), \quad s\geq1,\nonumber
\end{eqnarray}
where the $(1+n)th$-order cumulant of semigroups of operators (\ref{Sspher}) of hard spheres
with inelastic collisions is defined by the formula \cite{GerRS}
\begin{eqnarray}\label{cumulantd}
   &&\hskip-19mm\mathfrak{A}_{1+n}(t,\{Y\setminus X\},X)\doteq
       \sum\limits_{\mathrm{P}:\,(\{Y\setminus X\},\,X)={\bigcup}_i X_i}(-1)^{\mathrm{|P|}-1}({\mathrm{|P|}-1})!
       \prod_{X_i\subset \mathrm{P}}S_{|\theta(X_i)|}(t,\theta(X_i)),
\end{eqnarray}
and we used the notations: $Y\equiv(1,\ldots,s), Z\equiv (j_1,\ldots,j_{n})\subset Y$,
the set $\{Y\setminus Z\}$ is a set consisting of one element
$Y\setminus Z=(1,\ldots,j_{1}-1,j_{1}+1,\ldots,j_{n}-1,j_{n}+1,\ldots,s)$, i.e. this
set is a connected subset of the partition $\mathrm{P}$ such that $|\mathrm{P}|=1$,
the mapping $\theta(\cdot)$ is a declusterization operator defined by the formula
$\theta(\{Y\setminus Z\})=Y\setminus Z$.

The simplest examples of marginal observables (\ref{sdh}) are given by the following expansions:
\begin{eqnarray*}
   &&B_{1}(t,x_1)=\mathfrak{A}_{1}(t,1)B_{1}^0(x_1),\\
   &&B_{2}(t,x_1,x_2)=\mathfrak{A}_{1}(t,\{1,2\})B_{2}^0(x_1,x_2)+
      \mathfrak{A}_{2}(t,1,2)(B_{1}^0(x_1)+B_{1}^0(x_2)).
\end{eqnarray*}

Under the condition $\gamma<e^{-1}$, for the sequence of marginal observables (\ref{sdh})
the estimate holds
\begin{eqnarray}\label{es}
   &&\big\|B(t)\big\|_{\mathcal{C}_{\gamma}}
      \leq e^2(1-\gamma e)^{-1}\big\|B(0)\big\|_{\mathcal{C}_{\gamma}}.
\end{eqnarray}

For initial data $B(0)=(B_{0},B_{1}^0,\ldots,B_{s}^0,\ldots)\in C_{\gamma}^0\subset C_{\gamma}$
of finite sequences of infinitely differentiable functions with compact supports, the sequence of
functions (\ref{sdh}) is a classical solution and for arbitrary initial data $B(0)\in C_{\gamma}$
it is a generalized solution.

We note that single-component sequences of marginal observables correspond to observables
of a certain structure; namely, the marginal observable $B^{(1)}=(0,b_{1}(x_1),0,\ldots)$
corresponds to the additive-type observable, and the marginal observable
$B^{(k)}=(0,\ldots,0,b_{k}(x_1,$ $\ldots,x_k),0,\ldots)$ corresponds to the $k$-ary-type
observable \cite{BGer}. If in capacity of initial data (\ref{dhi}) we consider the
additive-type marginal observable, then the structure of solution expansion (\ref{sdh})
is simplified and attains the form
\begin{eqnarray}\label{af}
    &&B_{s}^{(1)}(t,x_1,\ldots,x_s)=\mathfrak{A}_{s}(t,1,\ldots,s)\sum_{j=1}^s b_{1}(x_j),\quad s\geq 1.
\end{eqnarray}

We remark that expansion (\ref{sdh}) can be also represented in the form of the weak formulation
of the perturbation (iteration) series \cite{BGer} as a result of applying of analogs of the
Duhamel equation to cumulants of semigroups of operators (\ref{cumulantd}).

\subsection{A functional of mean values of marginal observables}
The mean value of the marginal observables $B(t)\in \mathcal{C}_{\gamma}$
in the initial state described by the sequence of marginal distribution functions
$F(0)=(1,F_{1}^{0},\ldots,F_{n}^{0},\ldots)\in L^{1}_{\alpha}=\oplus^{\infty}_{n=0}\alpha^n L^{1}_{n}$
is determined by the functional
\begin{eqnarray}\label{avmar-1}
   &&\hskip-15mm\big\langle B(t)\big|F(0)\big\rangle=\sum\limits_{s=0}^{\infty}\,
      \frac{1}{s!}\int_{(\mathbb{R}^{3}\times\mathbb{R}^{3})^{s}}
      dx_{1}\ldots dx_{s}\,B_{s}(t,x_1,\ldots,x_s)F_{s}^{0}(x_1,\ldots,x_s).
\end{eqnarray}
Owing to estimate (\ref{es}), functional (\ref{avmar-1}) exists under the condition that
$\gamma<e^{-1}$.

In particular, functional (\ref{avmar-1}) of mean values of the additive-type marginal
observables $B^{(1)}(0)=(0,B_{1}^{(1)}(0,x_1),0,\ldots)$  takes the form
\begin{eqnarray*}\label{avmar-11}
   &&\hskip-12mm\big\langle B^{(1)}(t)\big|F(0)\big\rangle=\big\langle
       B^{(1)}(0)\big|F(t)\big\rangle=\int_{\mathbb{R}^{3}\times\mathbb{R}^{3}}dx_{1}\,
       B_{1}^{(1)}(0,x_1)F_{1}(t,x_1),
\end{eqnarray*}
where the one-particle marginal distribution function $F_{1}(t,x_1)$ is determined by
the series \cite{GerRS}
\begin{eqnarray*}
    &&\hskip-12mm F_{1}(t,x_1)=\sum\limits_{n=0}^{\infty}\frac{1}{n!}
      \int_{(\mathbb{R}^3\times\mathbb{R}^3)^n}dx_2\ldots dx_{n+1}\,
      \mathfrak{A}_{1+n}^\ast(t)F_{1+n}^0(x_1,\ldots,x_{n+1}),\nonumber
\end{eqnarray*}
and the generating operator $\mathfrak{A}_{1+n}^\ast(t)$ of this series expansion is the
$(1+n)th$-order cumulant of adjoint semigroups of operators of hard spheres with inelastic
collisions, i.e.
\begin{eqnarray}\label{nLkymyl}
   &&\mathfrak{A}_{1+n}^\ast(t,1,\ldots,n+1)=
      \sum\limits_{\texttt{P}:\,(1,\ldots,n+1)={\bigcup\limits}_i X_i}
      (-1)^{|\texttt{P}|-1}(|\texttt{P}|-1)!\prod_{X_i\subset \texttt{P}}S_{|\theta(X_i)|}^\ast(t,\theta(X_i)),
\end{eqnarray}
where it was used notations accepted in formula (\ref{cumulantd}).

In the general case for mean values of marginal observables, the following equality is true:
\begin{eqnarray*}\label{eqos}
   &&\big\langle B(t)\big|F(0)\big\rangle=\big\langle B(0)\big|F(t)\big\rangle,
\end{eqnarray*}
where the sequence $F(t)=(1,F_{1}(t),\ldots,F_{s}(t),\ldots)$ is a solution of the Cauchy problem
of the BBGKY hierarchy of hard spheres with inelastic collisions \cite{P08}. This equality signifies
the equivalence of two pictures of the description of the evolution of hard spheres by means of the
BBGKY hierarchy and the dual BBGKY hierarchy (\ref{dh}).

Furthermore, we consider initial states of hard spheres specified by a one-particle marginal distribution
function, namely
\begin{eqnarray}\label{eq:Bog2_haos}
    &&F^{(c)}_s(x_1,\ldots,x_s)=\prod_{i=1}^{s}F_{1}^0(x_i)
       \mathcal{X}_{\mathbb{R}^{3s}\setminus \mathbb{W}_s},\quad s\geq1,
\end{eqnarray}
where $\mathcal{X}_{\mathbb{R}^{3s}\setminus \mathbb{W}_s}\equiv\mathcal{X}_s(q_1,\ldots,q_s)$ is a
characteristic function of allowed configurations $\mathbb{R}^{3s}\setminus \mathbb{W}_s$ of $s$ hard
spheres and $F_{1}^0\in L^{1}(\mathbb{R}^{3}\times\mathbb{R}^{3})$. Initial data (\ref{eq:Bog2_haos})
is intrinsic for the kinetic description of many-particle systems because in this case all possible
states are characterized by means of a one-particle marginal distribution function.

\section{The non-Markovian Enskog equation for granular gases}
In view of the fact that initial state is completely specified by a one-particle marginal
distribution function, the evolution of states can be described within the framework of the
sequence $F(t\mid F_{1}(t))=(1,F_1(t),F_2(t\mid F_{1}(t)),\ldots,F_s(t\mid F_{1}(t)),\ldots)$
of the marginal functionals of the state $F_s(t,x_1,\ldots,x_s\mid F_{1}(t)),\,s\geq2$, which
are explicitly defined with respect to the solution $F_1(t,x_1)$ of the kinetic equation. We
refer to such a kinetic equation of inelastically interacting hard spheres as the non-Markovian
Enskog kinetic equation for granular gases.

\subsection{A description of the collective behavior of granular gases by means of kinetic equations}
In the case of initial states (\ref{eq:Bog2_haos}), the dual picture to the Heisenberg picture of the
evolution of a system of hard spheres with inelastic
collisions described in terms of the
dual BBGKY hierarchy (\ref{dh}) for marginal observables is the evolution of states described within
the framework of the non-Markovian Enskog kinetic equation and a sequence of explicitly defined
functionals of the solution of this kinetic equation.

In fact, for mean value functional (\ref{avmar-1}), the following equality holds:
\begin{eqnarray}\label{w}
    &&\big\langle B(t)\big|F^c\big\rangle=\big\langle B(0)\big|F(t\mid F_{1}(t))\big\rangle,
\end{eqnarray}
where $F^{(c)}=(1,F_{1}^{(c)},\ldots,F^{(c)}_s,\ldots)$ is the sequence of initial
marginal distribution functions (\ref{eq:Bog2_haos}), and the sequence
$F(t\mid F_{1}(t))=\big(1,F_1(t),F_2(t\mid F_{1}(t)),\ldots,F_s(t\mid F_{1}(t))\big)$
is a sequence of the marginal functional of the state $F_{s}(t,x_1,\ldots,x_s \mid F_{1}(t))$
represented by the series expansion over the products with respect to the one-particle
distribution function $F_{1}(t)$:
\begin{eqnarray}\label{f}
   &&\hskip-12mm F_{s}(t,x_1,\ldots,x_s\mid F_{1}(t))\doteq\\
   && \doteq\sum _{n=0}^{\infty }\frac{1}{n!}\,\int_{(\mathbb{R}^{3}\times\mathbb{R}^{3})^{n}}
      dx_{s+1}\ldots dx_{s+n}\,\mathfrak{V}_{1+n}(t,\{Y\},X\setminus Y)
      \prod_{i=1}^{s+n}F_{1}(t,x_i),\quad s\geq 2.\nonumber
\end{eqnarray}
In series (\ref{f}) we used the notations $Y\equiv(1,\ldots,s), X\equiv(1,\ldots,s+n)$;
and the $(n+1)th$-order generating operator $\mathfrak{V}_{1+n}(t)$, is defined as follows \cite{GG}:
\begin{eqnarray}\label{skrrn}
    &&\hskip-12mm \mathfrak{V}_{1+n}(t,\{Y\},X\setminus Y)\doteq\sum_{k=0}^{n}(-1)^k\,\sum_{m_1=1}^{n}\ldots
       \sum_{m_k=1}^{n-m_1-\ldots-m_{k-1}}\frac{n!}{(n-m_1-\ldots-m_k)!}\\
    &&\hskip-8mm \times\widehat{\mathfrak{A}}_{1+n-m_1-\ldots-m_k}(t,\{Y\},s+1,
       \ldots,s+n-m_1-\ldots-m_k)\prod_{j=1}^k\,\sum_{k_2^j=0}^{m_j}\ldots\nonumber\\
    &&\hskip-8mm \sum_{k^j_{n-m_1-\ldots-m_j+s}=0}^{k^j_{n-m_1-\ldots-m_j+s-1}}\,\prod_{i_j=1}^{s+n-m_1-\ldots-m_j}
       \frac{1}{(k^j_{n-m_1-\ldots-m_j+s+1-i_j}-k^j_{n-m_1-\ldots-m_j+s+2-i_j})!}\nonumber\\
    &&\hskip-8mm \times\widehat{\mathfrak{A}}_{1+k^j_{n-m_1-\ldots-m_j+s+1-i_j}-k^j_{n-m_1-\ldots-m_j+s+2-i_j}}(t,
       i_{j},s+n-m_1-\ldots-m_j+1\nonumber \\
    &&\hskip-8 mm+k^j_{s+n-m_1-\ldots-m_j+2-i_j},\ldots,
       s+n-m_1-\ldots-m_j+k^j_{s+n-m_1-\ldots-m_j+1-i_j}),\quad n\geq0,\nonumber
\end{eqnarray}
where it means that $k^j_1\equiv m_j,\,k^j_{n-m_1-\ldots-m_j+s+1}\equiv 0$, and
we denote the $(1+n)th$-order scattering cumulant by the operator $\widehat{\mathfrak{A}}_{1+n}(t)$:
\begin{eqnarray}\label{scacu}
   &&\hskip-8mm\widehat{\mathfrak{A}}_{1+n}(t,\{Y\},X\setminus Y)
      \doteq\mathfrak{A}_{1+n}^\ast(t,\{Y\},X\setminus Y)
      \mathcal{X}_{\mathbb{R}^{3(s+n)}\setminus \mathbb{W}_{s+n}}
      \prod_{i=1}^{s+n}\mathfrak{A}_{1}^\ast(t,i)^{-1},
\end{eqnarray}
and the operator $\mathfrak{A}_{1+n}^\ast(t)$ is the $(1+n)th$-order cumulant of adjoint semigroups
of hard spheres with inelastic collisions (\ref{nLkymyl}). We give several examples of expansions (\ref{skrrn}):
\begin{eqnarray*}
   &&\mathfrak{V}_{1}(t,\{Y\})=\widehat{\mathfrak{A}}_{1}(t,\{Y\})\doteq
       S_s^\ast(t,1,\ldots,s)\mathcal{X}_{\mathbb{R}^{3s}\setminus \mathbb{W}_{s}}
       \prod_{i=1}^{s}S_1^\ast(t,i)^{-1},\\
   &&\mathfrak{V}_{2}(t,\{Y\},s+1)=\widehat{\mathfrak{A}}_{2}(t,\{Y\},s+1)-
       \widehat{\mathfrak{A}}_{1}(t,\{Y\})\sum_{i_1=1}^s\widehat{\mathfrak{A}}_{2}(t,i_1,s+1).\nonumber
\end{eqnarray*}

We emphasize that, in fact, functionals (\ref{f}) characterize the correlations generated by dynamics
of a hard sphere system with inelastic collisions.

If $\|F_{1}(t)\|_{L^{1}(\mathbb{R}^{3}\times\mathbb{R}^{3})}<e^{-(3s+2)}$, then for arbitrary $t\geq0$
series (\ref{f}) converges in the norm of the space $L^{1}_{s}$.

The second element of the sequence $F(t\mid F_{1}(t))$, i.e. the one-particle marginal distribution
function $F_{1}(t)$, is determined by the following series expansion:
\begin{eqnarray}\label{F(t)1}
    &&\hskip-12mm F_{1}(t,x_1)=\sum\limits_{n=0}^{\infty}\frac{1}{n!}
      \int_{(\mathbb{R}^3\times\mathbb{R}^3)^n}dx_2\ldots dx_{n+1}\,\mathfrak{A}_{1+n}^\ast(t)
      \mathcal{X}_{\mathbb{R}^{3(1+n)}\setminus \mathbb{W}_{1+n}}\prod_{i=1}^{n+1}F_{1}^0(x_i),
\end{eqnarray}
where the generating operator $\mathfrak{A}_{1+n}^\ast(t)\equiv\mathfrak{A}_{1+n}^\ast(t,1,\ldots,n+1)$
is the $(1+n)th$-order cumulant of adjoint semigroups of hard spheres with inelastic
collisions defined by expansion (\ref{nLkymyl}).

For $t\geq 0$, the one-particle marginal distribution function (\ref{F(t)1}) is a solution
of the following Cauchy problem of the non-Markovian Enskog kinetic equation
\begin{eqnarray}
 \label{gke1}
   &&\hskip-12mm\frac{\partial}{\partial t}F_{1}(t,x_1)=
      -\langle p_1,\frac{\partial}{\partial q_1}\rangle F_{1}(t,x_1)+\\
   &&+\sigma^2\int_{\mathbb{R}^3\times\mathbb{S}^2_+}d p_2 d\eta
      \langle\eta,(p_1-p_2)\rangle\Big(\frac{1}{(1-2\varepsilon)^{2}}\,
      F_2(t,q_1,p_1^{\diamond},q_1-\sigma\eta,p_2^{\diamond}\mid F_{1}(t))-\nonumber\\
   &&-F_2(t,x_1,q_1+\sigma\eta,p_2\mid F_{1}(t))\Big),\nonumber\\ \nonumber\\
 \label{gkei}
   &&\hskip-12mm F_{1}(t)|_{t=0}= F_{1}^0,
\end{eqnarray}
where the collision integral is determined by the marginal functional of the state (\ref{f})
in the case of $s=2$, and the expressions $p_1^{\diamond}$ and $p_2^{\diamond}$ are the
pre-collision momenta of hard spheres with inelastic collisions (\ref{scol}), i.e. solutions
of equations (\ref{col}).

Thus, if initial states are specified by a one-particle marginal distribution function on allowed
configurations, then the evolution of marginal observables governed by the dual BBGKY hierarchy
(\ref{dh}) can be also described within the framework of the non-Markovian kinetic equation
(\ref{gke1}) and a sequence of marginal functionals of the state (\ref{f}). In other words, for
mentioned initial states, the evolution of all possible states of a hard sphere system with
inelastic collisions at an arbitrary moment of time can be described within the framework of a
one-particle distribution function without any approximations.

\subsection{The proof of the main results}
We establish the validity of equality (\ref{w}) for mean value functional (\ref{avmar-1}).

In the particular case of initial data (\ref{dhi}) specified by the $s$-ary marginal observable $s\geq2$,
i.e. the marginal observable $B^{(s)}(0)=(0,\ldots,0,b_{s},0,\ldots)$, equality (\ref{w})
takes the form
\begin{eqnarray}\label{avmar-12}
   &&\hskip-12mm\big\langle B^{(s)}(t)\big|F^{c}\big\rangle=
      \big\langle B^{(s)}(0)\big|F_s(t\mid F_{1}(t))\big\rangle=\\
   &&\hskip+12mm=\frac{1}{s!}\int_{(\mathbb{R}^{3}\times\mathbb{R}^{3})^{s}}
      dx_{1}\ldots dx_{s}\,b_{s}(x_1,\ldots,x_s)F_s(t,x_1,\ldots,x_s\mid F_{1}(t)), \quad s\geq2,\nonumber
\end{eqnarray}
where the marginal functional of the state $F_{s}(t,x_1,\ldots,x_s \mid F_{1}(t))$ is represented
by series expansion (\ref{f}).

To verify the validity of equality (\ref{avmar-12}), we transform the functional
$\big\langle B^{(s)}(t)\big|F^{c}\big\rangle$ to the form
\begin{eqnarray}\label{avmar-12t}
   &&\hskip-9mm\big\langle B^{(s)}(t)\big|F^{c}\big\rangle=\sum\limits_{n=0}^{\infty}\frac{1}{n!}
      \int_{(\mathbb{R}^{3}\times\mathbb{R}^{3})^{n}}dx_{1}\ldots dx_{n}\frac{1}{(n-s)!}\sum_{j_1\neq\ldots\neq j_{n-s}=1}^n
      \mathfrak{A}_{1+n-s}\big(t,\{1,\ldots,j_{1}-1,\\
    &&j_{1}+1,\ldots,j_{n-s}-1,j_{n-s}+1,\ldots,s\},j_1,\ldots,j_{n-s}\big)b_s(x_1,\ldots,x_s)
      \prod_{i=1}^{n}F_{1}^0(i)\mathcal{X}_{\mathbb{R}^{3n}\setminus \mathbb{W}_{n}}\nonumber\\
   &&=\frac{1}{s!}\int_{(\mathbb{R}^{3}\times\mathbb{R}^{3})^{s}}
      dx_{1}\ldots dx_{s}\,b_{s}(x_1,\ldots,x_s)\nonumber\\
   &&\times\sum\limits_{n=0}^{\infty}\frac{1}{n!}
      \int_{(\mathbb{R}^{3}\times\mathbb{R}^{3})^{n}}dx_{s+1}\ldots dx_{s+n}\,\mathfrak{A}_{1+n}^\ast(t,\{Y\},X\setminus Y)
      \prod_{i=1}^{s+n}F_{1}^0(i)\mathcal{X}_{\mathbb{R}^{3(s+n)}\setminus \mathbb{W}_{s+n}},\nonumber
\end{eqnarray}
where we used notations accepted in formula (\ref{f}), and the operator
$\mathfrak{A}_{1+n}^\ast(t,\{Y\},X\setminus Y)$ is the $(1+n)th$-order
cumulant of adjoint semigroups of hard spheres with inelastic collisions.
For $F_1^{0}\in L^{1}(\mathbb{R}\times\mathbb{R})$ and $b_{s}\in\mathcal{C}_s$
obtained functional (\ref{avmar-12t}) exists under the condition that
$\|F_1^0\|_{L^{1}(\mathbb{R}\times\mathbb{R})}<e^{-1}$.

Then we expand the cumulants $\mathfrak{A}_{1+n}^\ast(t),\,n\geq0,$ of adjoint semigroups
of hard spheres in functional (\ref{avmar-12t}) over the new evolution operators
$\mathfrak{V}_{1+n}(t),\,n\geq0,$ into the kinetic cluster expansions \cite{GG}:
\begin{eqnarray}\label{rrrl2}
   &&\hskip-12mm\mathfrak{A}_{1+n}^\ast(t,\{Y\},s+1,\ldots,s+n)\mathfrak{I}_{s+n}(1,\ldots,s+n)=\\ \nonumber
   &&\hskip-8mm=\sum_{k_1=0}^{n}\frac{n!}{(n-k_1)!k_1!}\,
     \mathfrak{V}_{1+n-k_1}(t,\{Y\},s+1,\ldots,s+n-k_1)\sum_{k_2=0}^{k_1}\frac{k_1!}{k_2!(k_1-k_2)!}\ldots\\ \nonumber
   &&\hskip-8mm\sum_{k_{n-k_1+s}=0}^{k_{n-k_1+s-1}}
     \frac{k_{n-k_1+s-1}!}{k_{n-k_1+s}!(k_{n-k_1+s-1}-k_{n-k_1+s})!}
     \prod_{i=1}^{s+n-k_1}{\mathfrak{A}}_{1+k_{n-k_1+s+1-i}-k_{n-k_1+s+2-i}}^\ast(t,i, \\ \nonumber
   &&\hskip-8mm s+n-k_1+1+k_{s+n-k_1+2-i},\ldots,s+n-k_1+k_{s+n-k_1+1-i})
     \mathfrak{I}_{1+k_{n-k_1+s+1-i}-k_{n-k_1+s+2-i}}(i,\nonumber\\
   &&\hskip-8mms+n-k_1+1+k_{s+n-k_1+2-i},\ldots,s+n-k_1+k_{s+n-k_1+1-i}),\quad n\geq0,\nonumber
\end{eqnarray}
where the following convention is assumed: $k_{s+1}\equiv 0$, and the operator $\mathfrak{I}_{s+n}$
is defined by the formula
\begin{eqnarray*}
   &&\mathfrak{I}_{s+n}(1,\ldots,s+n)f_{s+n}\doteq \mathcal{X}_{\mathbb{R}^{3(s+n)}\setminus\mathbb{W}_{s+n}}f_{s+n},
\end{eqnarray*}
where $\mathcal{X}_{\mathbb{R}^{3(s+n)}\setminus \mathbb{W}_{s+n}}$ is a characteristic function
of allowed configurations $\mathbb{R}^{3(s+n)}\setminus \mathbb{W}_{s+n}$ of a system of $s+n$
hard spheres.

We give several examples of recurrence relations (\ref{rrrl2}) in terms of scattering cumulants
(\ref{scacu}). Acting on both sides of equality (\ref{rrrl2}) by the evolution operators
$\prod_{i=1}^{s+n}\mathfrak{A}_{1}^\ast(t,i)^{-1}$, we obtain
\begin{eqnarray*}
   &&\widehat{\mathfrak{A}}_{1}(t,\{Y\})=\mathfrak{V}_{1}(t,\{Y\}),\\
   &&\widehat{\mathfrak{A}}_{2}(t,\{Y\},s+1)=\mathfrak{V}_{2}(t,\{Y\},s+1)+
       \mathfrak{V}_{1}(t,\{Y\})\sum_{i_1=1}^s \widehat{\mathfrak{A}}_{2}(t,i_1,s+1),\\
   &&\widehat{\mathfrak{A}}_{3}(t,\{Y\},s+1,s+2)=\mathfrak{V}_{3}(t,\{Y\},s+1,s+2)+\\
   &&\hskip+7mm +2!\mathfrak{V}_{2}(t,\{Y\},s+1)\sum_{i_1=1}^{s+1}\widehat{\mathfrak{A}}_{2}(t,i_1,s+2)
       +\mathfrak{V}_{1}(t,\{Y\})\big(\sum_{i_1=1}^s\widehat{\mathfrak{A}}_{3}(t,i_1,s+1,s+2)+\\
   &&\hskip+7mm +2!\sum_{1=i_1<i_2}^s\widehat{\mathfrak{A}}_{2}(t,i_1,s+1)\widehat{\mathfrak{A}}_{2}(t,i_2,s+2)\big),
\end{eqnarray*}
where the operator $\widehat{\mathfrak{A}}_{1+n}(t)$ is the $(1+n)th$-order scattering cumulant (\ref{scacu}).

We note that solutions of recurrence relations (\ref{rrrl2}) are given by expansions (\ref{skrrn}).

As a result of the application of kinetic cluster expansions (\ref{rrrl2}) the following equality is true
\begin{eqnarray*}
   &&\hskip-17mm\sum\limits_{n=0}^{\infty}\frac{1}{n!}
       \int_{(\mathbb{R}^{3}\times\mathbb{R}^{3})^{n}}dx_{s+1}\ldots dx_{s+n}
       \mathfrak{A}_{1+n}^\ast(t,\{Y\},X\setminus Y)\mathcal{X}_{\mathbb{R}^{3(s+n)}\setminus \mathbb{W}_{s+n}}
       \prod_{i=1}^{s+n}F_1^0(x_i)=\\
   &&=\sum\limits_{n=0}^{\infty}\frac{1}{n!}
       \int_{(\mathbb{R}^{3}\times\mathbb{R}^{3})^{n}}dx_{s+1}\ldots dx_{s+n}
       \mathfrak{V}_{1+n}(t,\{Y\},X\setminus Y)\prod_{i=1}^{s+n}F_{1}(t,x_i),
\end{eqnarray*}
where the $(n+1)th$-order generating evolution operator $\mathfrak{V}_{1+n}(t)$ is a solution of
recurrence relations (\ref{rrrl2}); i.e. it is determined by formula (\ref{skrrn}) and the function
$F_{1}(t)$ is represented by series expansion (\ref{F(t)1}).

Indeed, representing series over the summation index $n$ and the sum over the summation index $n_1$ in
functional (\ref{avmar-12t}) as a two-fold series, we derive
\begin{eqnarray*}
   &&\hskip-12mm\sum\limits_{n=0}^{\infty}\frac{1}{n!}\int\limits_{(\mathbb{R}^{3}\times\mathbb{R}^{3})^{n}}
      dx_{s+1}\ldots dx_{s+n}\,\mathfrak{A}_{1+n}^\ast(t,\{Y\},X\setminus Y)
      \mathcal{X}_{\mathbb{R}^{3(s+n)}\setminus \mathbb{W}_{s+n}}\prod_{i=1}^{s+n}F_{1}^0(x_i)=\\
   &&\hskip-5mm=\sum\limits_{n=0}^{\infty}\frac{1}{n!}\int_{(\mathbb{R}^{3}\times\mathbb{R}^{3})^n}dx_{s+1}\ldots
      dx_{s+n}\,\mathfrak{V}_{1+n}(t,\{Y\},X\setminus Y)\sum_{k_1=0}^{\infty}\sum_{k_2=0}^{k_1}\ldots\\
   &&\hskip-5mm\ldots\sum_{k_{n+s}=0}^{k_{n+s-1}}\frac{1}{k_{n+s}!(k_{n+s-1}-k_{n+s})!\ldots(k_1-k_2)!}
      \int_{(\mathbb{R}^{3}\times\mathbb{R}^{3})^{k_1}}dx_{n+s+1}\ldots\\
   &&\hskip-5mm \ldots dx_{n+s+k_1}\prod_{i=1}^{n+s}{\mathfrak{A}}_{1+k_{n+s+1-i}-k_{n+s+2-i}}^\ast
      (t,i,n+s+1+k_{n+s+2-i},\ldots,n+s+\\
   &&\hskip-5mm+k_{n+s+1-i})\mathcal{X}_{1+k_{n+s+1-i}-k_{n+s+2-i}}(q_i,q_{n+s+1+k_{n+s+2-i}},
      \ldots,q_{n+s+k_{n+s+1-i}})\prod_{j=1}^{n+s+k_1}F_1^0(x_j),
\end{eqnarray*}
where we used notations accepted above. According to the validity for series (\ref{F(t)1})
of the product formula
\begin{eqnarray}\label{Pr}
   &&\hskip-12mm\prod_{i=1}^{n+s}F_1(t,x_i)=\sum_{k_1=0}^{\infty}\sum_{k_2=0}^{k_1}\ldots
     \sum_{k_{n+s}=0}^{k_{n+s-1}}\frac{1}{k_{n+s}!(k_{n+s-1}-k_{n+s})!\ldots(k_1-k_2)!}\times \\
   &&\hskip-7mm\times\int_{(\mathbb{R}^{3}\times\mathbb{R}^{3})^{k_1}}dx_{n+s+1}\ldots dx_{n+s+k_1}
     \prod_{i=1}^{n+s}{\mathfrak{A}}_{1+k_{n+s+1-i}-k_{n+s+2-i}}^\ast(t,i,n+3+k_{n+4-i},\ldots,\nonumber\\
    &&\hskip-7mmn+2+k_{n+3-i})\mathcal{X}_{1+k_{n+s+1-i}-k_{n+s+2-i}}(q_i,
     q_{n+s+1+k_{n+4-i}},\ldots,q_{n+s+k_{n+3-i}})\prod_{j=1}^{n+s+k_1}F_1^0(x_j),\nonumber
\end{eqnarray}
in the obtained series expansion the series over the index $k_1$ can be expressed in terms of one-particle
distribution function (\ref{F(t)1}).

Thus, equality (\ref{avmar-12}) is valid.

In the case of initial data (\ref{dhi}) specified by the additive-type marginal observables, i.e.
$B^{(1)}(0)=(0,b_{1},0,\ldots)$, according to solution expansion (\ref{af}), equality (\ref{w})
takes the form
\begin{eqnarray}\label{avmar-11}
   &&\big\langle B^{(1)}(t)\big|F^{c}\big\rangle=
      \int_{\mathbb{R}^{3}\times\mathbb{R}^{3}}dx_{1}\,b_{1}(x_1)F_{1}(t,x_1),
\end{eqnarray}
where the one-particle marginal distribution function $F_{1}(t)$ is determined by series (\ref{F(t)1}).
This equality is proven in a similar manner to the first step of the proof of equality (\ref{avmar-12t}).

The validity of equality (\ref{w}) in case of the general type of marginal observables is proven
in much the same way as the validity of equalities (\ref{avmar-12}) and (\ref{avmar-11}).

Thus, we established the validity of statement (\ref{w}).

\subsection{The derivation of the non-Markovian Enskog kinetic equation with inelastic collisions}
Next, we establish that the one-particle marginal distribution function defined by series
expansion (\ref{F(t)1}) is governed by the non-Markovian Enskog kinetic equation (\ref{gke1}).

In view of the validity of the following equalities for cumulants
of adjoint semigroups of hard spheres with inelastic collisions in the sense of the norm convergence on the space
$L^{1}(\mathbb{R}^3\times\mathbb{R}^3)$
\begin{eqnarray*}
   &&\lim\limits_{t\rightarrow 0}\frac{1}{t}\mathfrak{A}_{1}^{\ast}(t,1)f_{1}(x_1)
     =\mathcal{L}^{\ast}_1(1)f_{1}(x_1),\\
   &&\lim\limits_{t\rightarrow 0}\frac{1}{t}\int_{\mathbb{R}^3\times\mathbb{R}^3}
     dx_2\mathfrak{A}_{2}^{\ast}(t,1,2)f_{2}(x_1,x_2)
     =\int_{\mathbb{R}^3\times \mathbb{R}^3}dx_2\mathcal{L}^{\ast}_{\mathrm{int}}(1,2)f_{2}(x_1,x_2),\\
   &&\lim\limits_{t\rightarrow 0}\frac{1}{t}
     \int_{\mathbb{R}^{3n}\times\mathbb{R}^{3n}}
     dx_2\ldots dx_{n+1}\,\mathfrak{A}_{1+n}^{\ast}(t,1,\ldots,n+1)f_{n+1}=0,\quad n\geq2,
\end{eqnarray*}
where the operators $\mathcal{L}^\ast_1(1)$ and $\mathcal{L}^\ast_{\mathrm{int}}(1,2)$ are given
by formulas (\ref{Lj}) and (\ref{aLint}) respectively, as a result of the differentiation over
the time variable of function (\ref{F(t)1}) in the sense of pointwise convergence on the space
$L^{1}(\mathbb{R}^3\times\mathbb{R}^3)$ we obtain
\begin{eqnarray}\label{de}
  &&\hskip-12mm\frac{\partial}{\partial t}F_{1}(t,x_1)=
     -\langle p_1,\frac {\partial}{\partial q_1}\rangle F_{1}(t,x_1)+
     \int_{\mathbb{R}^3\times\mathbb{R}^3}dx_2\mathcal{L}^{\ast}_{\mathrm{int}}(1,2)
     \sum\limits_{n=0}^{\infty}\frac{1}{n!}\times\\
  &&\hskip-5mm\times\int_{(\mathbb{R}^{3}\times\mathbb{R}^{3})^{n}}
     dx_{3}\ldots dx_{n+2}\,\mathfrak{A}_{1+n}^{\ast}(t,\{1,2\},3,\ldots,n+2)
     \prod_{i=1}^{n+2}F_1^0(x_i)\mathcal{X}_{\mathbb{R}^{3(n+2)}\setminus \mathbb{W}_{n+2}}.\nonumber
\end{eqnarray}
To represent the second term on the right-hand side in this equality in terms of one-particle
distribution function (\ref{F(t)1}), we expand cumulants $\mathfrak{A}_{1+n}^{\ast}(t,\{1,2\},3,\ldots,n+2)$
into kinetic cluster expansions (\ref{rrrl2}) in the case of $s=2$. Then we transform the series over
the summation index $n$ and the sum over the index $k_1$ to the two-fold series. As a result, it holds
\begin{eqnarray}\label{scint}
  &&\hskip-15mm\sum\limits_{n=0}^{\infty}\frac{1}{n!}\int_{(\mathbb{R}^{3}\times\mathbb{R}^{3})^{n}}
     dx_{3}\ldots dx_{n+2}\,\mathfrak{A}_{1+n}^{\ast}(t,\{1,2\},3,\ldots,n+2)
     \mathcal{X}_{\mathbb{R}^{3(n+2)}\setminus \mathbb{W}_{n+2}}\prod_{i=1}^{n+2}F_1^0(x_i)=\\
  &&\hskip-9mm=\sum\limits_{n=0}^{\infty}\frac{1}{n!}\sum_{k_1=0}^{\infty}
     \int_{(\mathbb{R}^{3}\times\mathbb{R}^{3})^{n+k_1}}dx_{3}\ldots dx_{n+2+k_1}
     \mathfrak{V}_{1+n}(t,\{1,2\},3,\ldots,n+2)\sum_{k_2=0}^{k_1}\ldots\nonumber\\
  &&\hskip-9mm\ldots\sum_{k_{n+2}=0}^{k_{n+1}}\frac{1}{k_{n+2}!(k_{n+1}-k_{n+2})!\ldots(k_1-k_2)!}
     \prod_{i=1}^{n+2}{\mathfrak{A}}_{1+k_{n+3-i}-k_{n+4-i}}^{\ast}(t,i,n+3+k_{n+4-i},\nonumber \\
  &&\hskip-9mm\ldots,n+2+k_{n+3-i})\mathcal{X}_{1+k_{n+3-i}-k_{n+4-i}}(q_i,q_{n+3+k_{n+4-i}},
     \ldots,q_{n+2+k_{n+3-i}})\prod_{j=1}^{n+2+k_1}F_1^0(x_j). \nonumber
\end{eqnarray}
According to equality (\ref{Pr}) in the case of $s=2$, the last series in equality (\ref{scint})
can be expressed in terms of series expansion (\ref{F(t)1}) for a one-particle marginal distribution
function.

We treat the constructed identity for a one-particle distribution function as the kinetic equation
for a hard sphere system with inelastic collisions. We refer to this evolution equation as the non-Markovian
Enskog kinetic equation.

Hence, for the additive-type marginal observables, the non-Markovian Enskog kinetic equation (\ref{gke1})
is dual with respect to bilinear form (\ref{avmar-1}) to the dual BBGKY hierarchy for hard spheres with
inelastic collisions (\ref{dh}).

\subsection{Some properties of the collision integral}
We represent the collision integral $\mathcal{I}_{E}$ of the non-Markovian Enskog equation
(\ref{gke1}) in the form of an expansion with respect to the Boltzman-Enskog collision integral
for hard spheres with inelastic collisions
\begin{eqnarray*}
  &&\hskip-18mm\mathcal{I}_{BE}\equiv\sigma^2\int_{\mathbb{R}^3\times\mathbb{S}^2_+}d p_2 d\eta
     \langle\eta,(p_1-p_2)\rangle
     \big(\frac{1}{(1-2\varepsilon)^{2}}\,F_1(t,q_1,p^\diamond_1)F_1(t,q_1-\sigma\eta,p^\diamond_2)-\\
  &&-F_1(t,x_1)F_1(q_1+\sigma\eta,p_2)\big),
\end{eqnarray*}
where the momenta $p_{1}^{\diamond},p_{2}^{\diamond}$ are determined by expressions (\ref{scol}).
We observe that such expansion of the collision integral $\mathcal{I}_{E}$ is expressed in terms
of the marginal correlation functional
\begin{eqnarray*}
    &&G_{2}\big(t,x_1,x_2\mid F_{1}(t)\big)=F_{2}\big(t,x_1,x_2\mid F_{1}(t)\big)-F_{1}(t,x_1)F_{1}(t,x_2).
\end{eqnarray*}

Indeed, in view of the validity of the equality
\begin{eqnarray*}
    &&(\widehat{\mathfrak{A}}_{1}(t,\{1^{\sharp},2^{\sharp}_{\pm}\})-I)
           F_{1}(t,q_1,p_1^{\sharp})F_{1}(t,q_1\pm\sigma\eta,p_2^{\sharp})=0,
\end{eqnarray*}
where we used notations adopted to the conventional notation of the Enskog collision integral \cite{GG},
we derive
\begin{eqnarray}\label{GEE}
  &&\hskip-15mm\mathcal{I}_{E}=\mathcal{I}_{BE}+\sum_{n=1}^{\infty}\mathcal{I}^{(n)}_{E}\equiv\\
  &&\hskip-15mm\equiv\mathcal{I}_{BE}+\sigma^2\sum_{n=1}^{\infty}\frac{1}{n!}
     \int_{\mathbb{R}^3\times\mathbb{S}^2_+}d p_2 d\eta
     \int_{(\mathbb{R}^{3}\times\mathbb{R}^{3})^{n}}dx_3\ldots dx_{n+2}\langle\eta,(p_1-p_2)\rangle\times\nonumber\\
  &&\hskip-8mm\big(\frac{1}{(1-2\varepsilon)^{2}}\,
     \mathfrak{V}_{1+n}(t,\{1^{\diamond},2^{\diamond}_{-}\},3,\ldots,n+2)
     F_1(t,q_1,p_1^{\diamond})F_1(t,q_1-\sigma\eta,p_2^{\diamond})\prod_{i=3}^{n+2}F_{1}(t,x_i)-\nonumber\\
  &&\hskip-8mm-\mathfrak{V}_{1+n}(t,\{1,2_{+}\},3,\ldots,n+2)F_1(t,x_1)
     F_1(t,q_1+\sigma\eta,p_2)\prod_{i=3}^{n+2}F_{1}(t,x_i)\big).\nonumber
\end{eqnarray}

We remark that the properties of a solution of kinetic equation (\ref{gke1}) for hard spheres with inelastic
collisions are determined by the properties of generating operators of expansion (\ref{GEE}) of the collision
integral.

For the abstract Cauchy problem (\ref{gke1}),(\ref{gkei}) of the non-Markovian Enskog kinetic equation
in the space of integrable functions $L^{1}(\mathbb{R}^3\times\mathbb{R}^3)$, the following statement
is true.

A global in time solution of the Cauchy problem (\ref{gke1}),(\ref{gkei}) of the non-Markovian Enskog
equation is represented by series expansion (\ref{F(t)1}).
If $\|F_1^0\|_{L^{1}(\mathbb{R}^3\times\mathbb{R}^3)}<e^{-10}(1+e^{-9})^{-1}$, then for
$F_1^0\in{L}^{1}_{0}(\mathbb{R}^3\times\mathbb{R}^3)$ function (\ref{F(t)1}) is a strong solution
and for an arbitrary initial data $F_1^{0}\in L^{1}(\mathbb{R}^3\times\mathbb{R}^3)$ it is a weak solution.

We also note, that on the basis of the derived non-Markovian Enskog equation (\ref{gke1}) we can
formulate the Markovian Enskog kinetic equation with inelastic collisions. The Markovian approximation
means that the generating operators of the collision integral $\mathcal{I}_{E}$ must be replaced on the
limit operators: $\mathfrak{V}_{1+n}(\{1,2\},3,\ldots,n+2)=
\lim_{\epsilon\rightarrow 0}\mathfrak{V}_{1+n}(\epsilon^{-1}t,\{1,2\},3,\ldots,n+2)$, where $\epsilon>0$
is the suitable scale parameter \cite{CGP97}.

\section{On the Boltzmann--Grad asymptotic behavior of one-dimensional granular gases}

It is well known that in the Boltzmann--Grad scaling limit \cite{GH},\cite{Sp80}, the dynamics of
a one-dimensional system of hard spheres with elastic collisions is trivial (a free motion or
the Knudsen flow) \cite{PG1}. However, as observed in paper \cite{NY1}, with this scaling
limit, the dynamics of inelastically interacting hard rods is not trivial and it is governed by the
Boltzmann kinetic equation for granular gases \cite{BCP2},\cite{T1}. In this section, the approach
to the rigorous derivation of Boltzmann-type equation for one-dimensional granular gases is outlined.
We remark that a one-dimensional system of hard spheres, i.e. hard rods, with inelastic collisions
exhibits the essential properties of granular gases in view that in a multidimensional case under
the inelastic collisions, only the normal component of relative velocities dissipates \cite{BCP2}.

\subsection{The Boltzmann--Grad limit of the non-Markovian Enskog equation with inelastic collisions}
We consider the Boltzmann--Grad asymptotic behavior of the non-Markovian Enskog equation solution
for a one-dimensional granular gas. In this case, for $t\geq0$ in dimensionless form, the Cauchy problem
(\ref{gke1}),(\ref{gkei}) has the form
\begin{eqnarray}\label{GE1}
  &&\hskip-12mm\frac{\partial}{\partial t}F_{1}(t,x_1)=
      -p_1\frac{\partial}{\partial q_1}F_{1}(t,x_1)+\\
  &&+\int_0^\infty dP\,P\big(\frac{1}{(1-2\varepsilon)^2}\,
      F_{2}(t,q_1,p_1^\diamond(p_1,P),q_1-\epsilon,p_{2}^\diamond(p_1,P)\mid F_{1}(t))-\nonumber\\
  &&-F_{2}(t,q_1,p_1,q_1-\epsilon,p_1+P\mid F_{1}(t))\big)+\nonumber\\
  &&+\int_0^\infty dP\,P \big(\frac{1}{(1-2\varepsilon)^2}\,F_{2}(t,q_1,\tilde p_1^\diamond(p_1,P),q_1+
      \epsilon,\tilde p_{2}^\diamond(p_1,P)\mid F_{1}(t))-\nonumber\\
  &&-F_{2}(t,q_1,p_1,q_1+\epsilon,p_1-P\mid F_{1}(t))\big),\nonumber\\ \nonumber\\
  \label{GKEi}
   &&\hskip-12mm F_{1}(t)|_{t=0}= F_{1}^{\epsilon,0},
\end{eqnarray}
where $\epsilon>0$ is a scaling parameter (the ratio of a hard sphere diameter (the length) $\sigma>0$ to the
mean free path), the collision integral is determined by the marginal functional of the state (\ref{f})
in the case of $s=2$, and the expressions
\begin{eqnarray*}
   &&p_{1}^\diamond(p_1,P)=p_1-P+\frac{\varepsilon}{2\varepsilon -1}\,P,\\
   &&p_{2}^\diamond(p_1,P)=p_1-\frac{\varepsilon}{2\varepsilon -1}\,P\nonumber
\end{eqnarray*}
and
\begin{eqnarray*}
   &&\tilde p_{1}^\diamond(p_1,P)=p_1+P-\frac{\varepsilon}{2\varepsilon -1}\,P,\\
   &&\tilde p_{2}^\diamond(p_1,P)=p_1+\frac{\varepsilon}{2\varepsilon -1}\,P,\nonumber
\end{eqnarray*}
are transformed pre-collision momenta in a one-dimensional space.

If initial one-particle marginal distribution functions satisfy the following condition:
\begin{eqnarray}\label{G_1}
    &&|F_{1}^{\epsilon,0}(x_1)|\leq Ce^{\textstyle-\frac{\beta}{2}{p^{2}_1}},
\end{eqnarray}
where $\textstyle{\beta}>0$ is a parameter and $C<\infty$ is some constant, then every term of
the series
\begin{eqnarray}\label{ske1}
  &&\hskip-9mm F_{1}^{\epsilon}(t,x_1)=\sum\limits_{n=0}^{\infty}\frac{1}{n!}
     \int_{(\mathbb{R}\times\mathbb{R})^n}dx_2\ldots dx_{n+1}\,\mathfrak{A}_{1+n}^\ast(t,1,\ldots,n+1)
     \prod_{i=1}^{n+1}F_{1}^{\epsilon,0}(x_i)\mathcal{X}_{\mathbb{R}^{(1+n)}\setminus \mathbb{W}_{1+n}}^{\epsilon},
\end{eqnarray}
exists, for finite time interval function (\ref{ske1}) is the uniformly convergent series with
respect to $x_1$ from arbitrary compact. Series expansion (\ref{ske1}) is represented a weak solution
of the Cauchy problem (\ref{GE1}),(\ref{GKEi}) of the non-Markovian Enskog equation.

The proof of this statement is based on analogs of the Duhamel equations for cumulants
$\mathfrak{A}_{1+n}^\ast(t),\,n\geq1,$ of semigroups of adjoint operators of hard spheres and
estimates established for the iteration series of the BBGKY hierarchy for hard spheres with elastic
collisions \cite{PG1},\cite{R10},\cite{G92}.

The following Boltzmann--Grad limit theorem is true.

If initial marginal one-particle distribution function $F_{1}^{\epsilon,0}$ satisfies condition
(\ref{G_1}) and in the sense of a weak convergence of the space of bounded functions there exists
the limit
\begin{eqnarray}\label{asumdin}
   &&\mathrm{w-}\lim\limits_{\epsilon\rightarrow 0}\big(F_{1}^{\epsilon,0}(x_1)-f_{1}^0(x_1)\big)=0,
\end{eqnarray}
then for finite time interval there exists the Boltzmann--Grad limit of solution (\ref{ske1}) of
the Cauchy problem (\ref{GE1}),(\ref{GKEi}) of the non-Markovian Enskog equation for granular gas
\begin{eqnarray}\label{asymt}
   &&\mathrm{w-}\lim\limits_{\epsilon\rightarrow 0}\big(F_{1}^{\epsilon}(t,x_1)-f_{1}(t,x_1)\big)=0,
\end{eqnarray}
where the limit one-particle marginal distribution function is defined by uniformly convergent
on arbitrary compact set series:
\begin{eqnarray}\label{viter}
  &&\hskip-7mm f_{1}(t,x_1)=\sum\limits_{n=0}^{\infty}\frac{1}{n!}
     \int_{(\mathbb{R}\times\mathbb{R})^n}dx_2\ldots dx_{n+1}\,\mathfrak{A}_{1+n}^0(t,1,\ldots,n+1)
     \prod_{i=1}^{n+1}f_{1}^0(x_i),
\end{eqnarray}
and the generating operator $\mathfrak{A}_{1+n}^0(t)$ is the $(n+1)th$-order
cumulant of adjoint semigroups $S_n^{\ast,0}(t)$ of point particles with inelastic collisions in
a one-dimensional space. An infinitesimal generator of the adjoint semigroups $S_n^{\ast,0}(t),\,n\geq1,$
is defined by the operator
\begin{eqnarray*}
    &&\hskip-12mm (\mathcal{L}_{n}^{\ast,0}f_{n})(x_{1},\ldots,x_{n})=
       -\sum_{j=1}^{n}p_j\,\frac{\partial}{\partial q_j} f_{n}(x_{1},\ldots,x_{n})+\\
    && +\sum_{j_{1}<j_{2}=1}^{n}|p_{j_{2}}-p_{j_{1}}|\big(\frac{1}{(1-2\varepsilon)^{2}}
       f_{n}(x_{1},\ldots,q_{j_{1}},p^{\diamond}_{j_{1}},\ldots,q_{j_{2}},p^{\diamond}_{j_{2}},\ldots,x_{n})-\\
    && -f_{n}(x_{1},\ldots,x_{n})\big)\delta(q_{j_{1}}-q_{j_{2}}),
\end{eqnarray*}
where the pre-collision momenta $p^\diamond_{j_{1}},\,p^\diamond_{j_{2}}$ are determined by
the following expressions:
\begin{eqnarray*}
  && p^{\diamond}_{j_{1}}=p_{j_{2}}+\frac{\varepsilon}{2\varepsilon -1}\,(p_{j_{1}}-p_{j_{2}}),\\
  && p_{j_{2}}^{\diamond}=p_{j_{1}}-\frac{\varepsilon}{2\varepsilon-1}\,(p_{j_{1}}-p_{j_{2}}).\nonumber
\end{eqnarray*}

If $f_{1}^0$ satisfies condition (\ref{G_1}), then for $t\geq 0$ the limit one-particle distribution
function represented by series expansion (\ref{viter}) is a weak solution of the Cauchy problem of the
Boltzmann-type kinetic equation of point particles with inelastic collisions in a one-dimensional space
\begin{eqnarray}\label{Bolz}
  &&\hskip-12mm \frac{\partial}{\partial t}f_1(t,q,p)=-p\,\frac{\partial}{\partial q}f_1(t,q,p)+
     \int_{-\infty}^{+\infty}d p_1\,|p-p_1|\times\\
  && \times\big(\frac{1}{(1-2\varepsilon)^2}\,f_1(t,q,p^\diamond)\,f_1(t,q,p^\diamond_1)
     -f_1(t,q,p)\,f_1(t,q,p_1)\big)+\sum_{n=1}^{\infty}\mathcal{I}^{(n)}_{0}.\nonumber
\end{eqnarray}
In kinetic equation (\ref{Bolz}) the remainder $\sum_{n=1}^{\infty}\mathcal{I}^{(n)}_{0}$ of the
collision integral of the kinetic equation is determined by the expressions $\mathcal{I}^{(n)}_{0}$,
which have the similar structure as $\mathcal{I}^{(n)}_{E}$ from series expansion (\ref{GEE})
\begin{eqnarray*}
  &&\hskip-18mm\mathcal{I}^{(n)}_{0}\equiv\frac{1}{n!}\int_0^\infty dP\,P\,
      \int_{(\mathbb{R}\times\mathbb{R})^{n}}dx_3\ldots dx_{n+2}\mathfrak{V}_{1+n}(t)\big(\frac{1}{(1-2\varepsilon)^2}
      F_1(t,q,p_{1}^\diamond(p,P))\times\nonumber\\
  &&\hskip-8mm\times F_1(t,q,p_{2}^\diamond(p,P))-F_1(t,q,p)F_1(t,q,p+P)\big)
      \prod_{i=3}^{n+2}F_{1}(t,x_i)+\nonumber\\
  &&\hskip-8mm+\int_0^\infty dP\,P\,\int_{(\mathbb{R}\times\mathbb{R})^{n}}dx_3\ldots dx_{n+2}
      \mathfrak{V}_{1+n}(t)\big(\frac{1}{(1-2\varepsilon)^2}
      F_1(t,q,\tilde p_{1}^\diamond(p,P))\times\nonumber\\
  &&\hskip-8mm\times F_1(t,q,\tilde p_{2}^\diamond(p,P))-F_1(t,q,p)F_1(t,q,p-P)\big)
      \prod_{i=3}^{n+2}F_{1}(t,x_i),\nonumber
\end{eqnarray*}
where the generating operators $\mathfrak{V}_{1+n}(t)\equiv\mathfrak{V}_{1+n}(t,\{1,2\},3,\ldots,n+2),\,n\geq0,$
are represented by expansions (\ref{skrrn}) with respect to the cumulants of scattering
operators of point hard rods with inelastic collisions in a one-dimensional space
\begin{eqnarray}\label{scat}
   &&\widehat{S}_{n}^{0}(t,1,\ldots,n)\doteq
       S_n^{\ast,0}(t,1,\ldots,s)\prod_{i=1}^{n}S_1^{\ast,0}(t,i)^{-1}.
\end{eqnarray}

In fact, the series expansions for the collision integral or solution (\ref{ske1}) of the
non-Markovian Enskog equation for a granular gas (\ref{GE1}) are represented as the power
series over the density, so that the terms $\mathcal{I}^{(n)}_{0},\,n\geq1,$ of the collision
integral of the Boltzmann-type kinetic equation (\ref{Bolz}) are corrections with respect to
the density to the Boltzmann collision integral for a one-dimensional granular gas, as claimed
previously \cite{BCP2},\cite{T1}.

We remark that since the scattering operator of point hard rods is an identity operator in
the limit of elastic collisions, i.e. in the limit $\varepsilon\rightarrow0$, the collision
integral of the Boltzmann-type kinetic equation (\ref{Bolz}) in a one-dimensional space is
equal to zero. In the quasi-elastic limit \cite{T1}, the limit one-particle marginal distribution
function satisfies the nonlinear friction kinetic equation for one-dimensional granular
gases \cite{NY1},\cite{T1}.

\subsection{On the propagation of a chaos in granular gases}
Taking into consideration the results (\ref{asymt}) on the Boltzmann--Grad asymptotic behavior
of the non-Markovian Enskog equation (\ref{gke1}), for marginal functionals of the state
(\ref{f}) in a one-dimensional space, the following statement is true.

For finite time interval in the sense of a weak convergence of the space of bounded functions
for marginal functionals of the state (\ref{f}), it holds
\begin{eqnarray*}
  &&\hskip-9mm\mathrm{w-}\lim\limits_{\epsilon\rightarrow 0}
      \big(F_{s}\big(t,x_1,\ldots,x_s\mid F_{1}^{\epsilon}(t)\big)
      -f_{s}\big(t,x_1,\ldots,x_s\mid f_{1}(t)\big)\big)=0,\quad s\geq2,
\end{eqnarray*}
where the limit marginal functionals $f_{s}\big(t\mid f_{1}(t)\big),\,s\geq2,$ with respect to
limit one-particle distribution function (\ref{viter}) are represented by the series expansion
\begin{eqnarray*}
   &&\hskip-7mm f_{s}(t,x_1,\ldots,x_s\mid f_{1}(t))\doteq\sum _{n=0}^{\infty }\frac{1}{n!}\,
      \int_{(\mathbb{R}^{3}\times\mathbb{R}^{3})^{n}} dx_{s+1}\ldots dx_{s+n}\,
      \mathfrak{V}_{1+n}^{0}(t,\{Y\},X\setminus Y)\prod_{i=1}^{s+n}f_{1}(t,x_i),
\end{eqnarray*}
where the generating operators $\mathfrak{V}_{1+n}^{0}(t),\,n\geq0,$ are represented by expansions
(\ref{skrrn}) over the cumulants of scattering operators (\ref{scat}) of point hard rods with
inelastic collisions.

In the case of a hard rod system with elastic collisions, the limit marginal functionals are
products of the limit one-particle distribution function, which describes the free motion of
point hard rods, which means the propagation of initial chaos.

Thus, in the Boltzmann--Grad scaling limit, solution (\ref{ske1}) of the non-Markovian Enskog
equation (\ref{GE1}) is governed by the Boltzmann-type equation (\ref{Bolz}) for one-dimensional
granular gases. The limit marginal functionals of the state are represented by the corresponding
series expansions with respect to the limit one-particle distribution function (\ref{viter}), which
describes how the initial chaos propagates in one-dimensional granular gases.

We also point out that in a multidimensional space, the Boltzmann--Grad asymptotic behavior of
the non-Markovian Enskog equation (\ref{gke1}) is similar to the Boltzmann--Grad asymptotic
behavior of a system of hard spheres with elastic collisions \cite{PG1}; namely, it is governed by
the Boltzmann-type equation for granular gases \cite{GG04} and the limit marginal functionals of
the state are represented by the products of its solution, which means the property of the propagation
of initial chaos.

\section{Conclusion and outlook}
The origin of the microscopic description of the evolution of observables of a hard
sphere system with inelastic collisions was considered. In case of initial states
(\ref{eq:Bog2_haos}) specified by a one-particle distribution function solution (\ref{sdh}),
the Cauchy problem of the dual BBGKY hierarchy for hard spheres with inelastic
collisions (\ref{dh}),(\ref{dhi}) and a solution of the Cauchy problem of the
non-Markovian Enskog equation for granular gases (\ref{gke1}),(\ref{gkei}) together
with marginal functionals of the state (\ref{f}), give two equivalent approaches to
the description of the evolution of a hard sphere system with inelastic collisions.
In fact, the rigorous justification of the Enskog kinetic equation for granular gases
is a consequence of the validity of equality (\ref{w}).

We note that the structure of the collision integral expansion (\ref{GEE}) of the
non-Markovian Enskog equation for granular gases (\ref{gke1}) is such that the first
term of this expansion is the Boltzmann--Enskog collision integral and the next terms describe
all possible correlations generated by hard sphere dynamics with inelastic collisions
and by the propagation of initial correlations connected with the forbidden configurations.

One of the advantages of the developed approach is the possibility to construct the kinetic
equations in scaling limits, involving correlations at initial time which can characterize
the condensed states of a hard sphere system with inelastic collisions.

We emphasize that the approach to the derivation of the Boltzmann equation with inelastic
collisions from underlying dynamics governed by the non-Markovian Enskog kinetic equation
for granular gases enables to construct the higher-order corrections to the Boltzmann--Grad
evolution of many hard spheres with inelastic collisions.

Finally, we remark that the developed approach is also related to the problem of a rigorous
derivation of the non-Markovian kinetic-type equations from underlaying many-particle dynamics,
which make it possible to describe the memory effects in granular gases.


\addcontentsline{toc}{section}{References}
{\small
\renewcommand{\refname}{References}

\end{document}